\documentclass[%
 reprint,
superscriptaddress,
 amsmath,amssymb,
 aps,
]{revtex4-2}

\usepackage{graphicx}
\usepackage{dcolumn}
\usepackage{bm}
\usepackage{natbib} 
\usepackage{makecell}
\usepackage{xcolor}
\usepackage{caption}

\begin{document}

\preprint{APS/123-QED}

\title{Nonlinear Wave Propagation in 1D Polycatenated Ring Chains}

\author{Xiaoxiao Xiong}
\thanks{These authors contributed equally to this work.}  
\affiliation{Division of Physics, Math, and Astronomy, California Institute of Technology, Pasadena, CA, USA}

\author{Reo Yanagi}
\thanks{These authors contributed equally to this work.} 
\author{Tingtao Zhou}
\author{Chiara Daraio}
\email{daraio@caltech.edu}  
\affiliation{Division of Engineering and Applied Science, California Institute of Technology, Pasadena, CA, USA}

\date{\today}

\begin{abstract}

We study the nonlinear wave dynamics of one-dimensional chains of polycatenated rings. These interlocked structures support amplitude-dependent nonlinear wave propagation driven by tensile activation and internal structural flexibility, unlike traditional granular crystals. Through dynamic impact experiments, finite-element modeling, and discrete-particle simulations of vertical chains pretensioned by gravity, we observe and explain nonlinear waves characterized by a compact leading wavefront followed by persistent trailing oscillations, which arise from energy partitioning into the rings' internal bending modes. Further, we demonstrate that the system's nonlinearity is not a fixed material constant. By altering the rings' geometric aspect ratio and contact angles, we can tune the effective contact exponent and the amplitude scaling of the wave speed. This work builds upon nonlinear wave propagation in classical granular crystals and establishes polycatenated systems as a highly tunable and designable platform to study and control nonlinear dynamics.

\end{abstract}

\maketitle
\def\thefootnote{*}\footnotetext{These authors contributed equally to this work}


\section{\label{sec:1}Introduction}

A classical ring chain, composed of a series of freely interlinked, closed loops, is one of the most ubiquitous structural motifs in everyday life. Despite its many applications, its response under high-rate impact harbors complex, largely uncharacterized nonlinear phenomena. Recently, this fundamental architectural concept has been scaled and formalized into a new class of mechanical metamaterials known as polycatenated architected materials (PAMs) \cite{structured-fab-2021, polycatenated_2025}. Constructed from interlinked cage- or ring-like particles assembled into lattices, PAMs serve as a bridge between unbound granular systems and rigid continuous structures. Because of their unique ability to dissipate or localize energy through complex internal deformations, they have been envisioned for advanced dynamic applications such as shock-absorbing materials \cite{teflon_beads_soliton_2005, energy_trapping_2006, optimal_granular_2009, Anomalous_wave_reflection_2005}. However, existing research has primarily focused on their quasi-static behavior; how mechanical waves actually propagate through these topologically constrained, interconnected systems remains an open question.

To analyze the dynamics of such chains, a natural point of comparison is the extensively studied field of strongly nonlinear wave propagation in one-dimensional granular media \cite{Nesterenko-book-2001,nesterenko_propagation_1984, lazaridi_observation_1985, Nesterenko_1994, sen_solitary_2008, sinkovits_nonlinear_1995, coste_solitary_1997, Tunability_2006, porter_highly_2008,ponson_nonlinear_2010, boechler_discrete_2010, teflon_beads_soliton_2005, ellipsoidal_2011, cylindrical_2012, hollow_2013, energy_trapping_2006, optimal_granular_2009, Anomalous_wave_reflection_2005, bifurcation_switch_2011}. Traditional granular chains are weakly compressed assemblies of particles with zero tensile strength. They exhibit a sonic vacuum regime in which linear acoustic wave propagation is suppressed, instead supporting the formation of a distinct type of solitary wave. Signatures of these solitary waves include a spatial width characterized by the nonlinear contact exponent and a strongly amplitude-dependent wave speed \cite{Nesterenko-book-2001}. Extensive research has demonstrated that these nonlinear dynamics can be deterministically tuned via initial prestress \cite{Tunability_2006}, periodicity \cite{porter_highly_2008,ponson_nonlinear_2010, boechler_discrete_2010}, material properties \cite{teflon_beads_soliton_2005}, and particle geometries \cite{ellipsoidal_2011, cylindrical_2012, hollow_2013}.

In this paper, we bridge the gap between classical granular physics and the dynamics of interlinked metamaterials by extending the exploration of nonlinear solitary waves to one-dimensional chains composed of polycatenated tori. Analogous to traditional discrete media that rely on a compression-only contact, a polycatenated chain requires tension to pull the interlocked rings into contact. However, instead of treating particles as quasi-rigid bodies governed purely by local Hertzian surface interactions, the structural flexibility of the rings introduces critical internal degrees of freedom. These bending modes partition wave energy and alter the propagating waveform, effectively shifting the system's nonlinearity from a fixed material property to a tunable and designable geometric parameter.

The remainder of this paper is structured as follows. In 
Section~\ref{sec: exp} we develop a mechanical model of the inter-ring 
interaction, treating each contact as a series combination of nonlinear 
Hertzian contact and linear ring-body bending, and reduce this hybrid 
response to a single-parameter power-law abstraction $F = K\delta^n$. 
Section~\ref{sec: wave prop} presents dynamical experiments on a 1D 
ring chain and compares the observed amplitude-dependent solitary waves 
and trailing oscillations against discrete-particle simulations of both 
the dual-spring and power-law frameworks. Section~\ref{sec: tunability} 
examines the geometric tunability of the effective nonlinear exponent $n$ 
as a function of ring aspect ratio $R/r$ and inter-ring contact angle 
$\theta$, and maps the resulting wave-speed scaling onto Nesterenko's 
long-wavelength theory. Section~\ref{sec: discussion} concludes by 
discussing polycatenated architected materials as a highly tunable and designable
platform for studying nonlinear dynamics.

\begin{figure*}
    \centering
    \includegraphics[width=\linewidth]{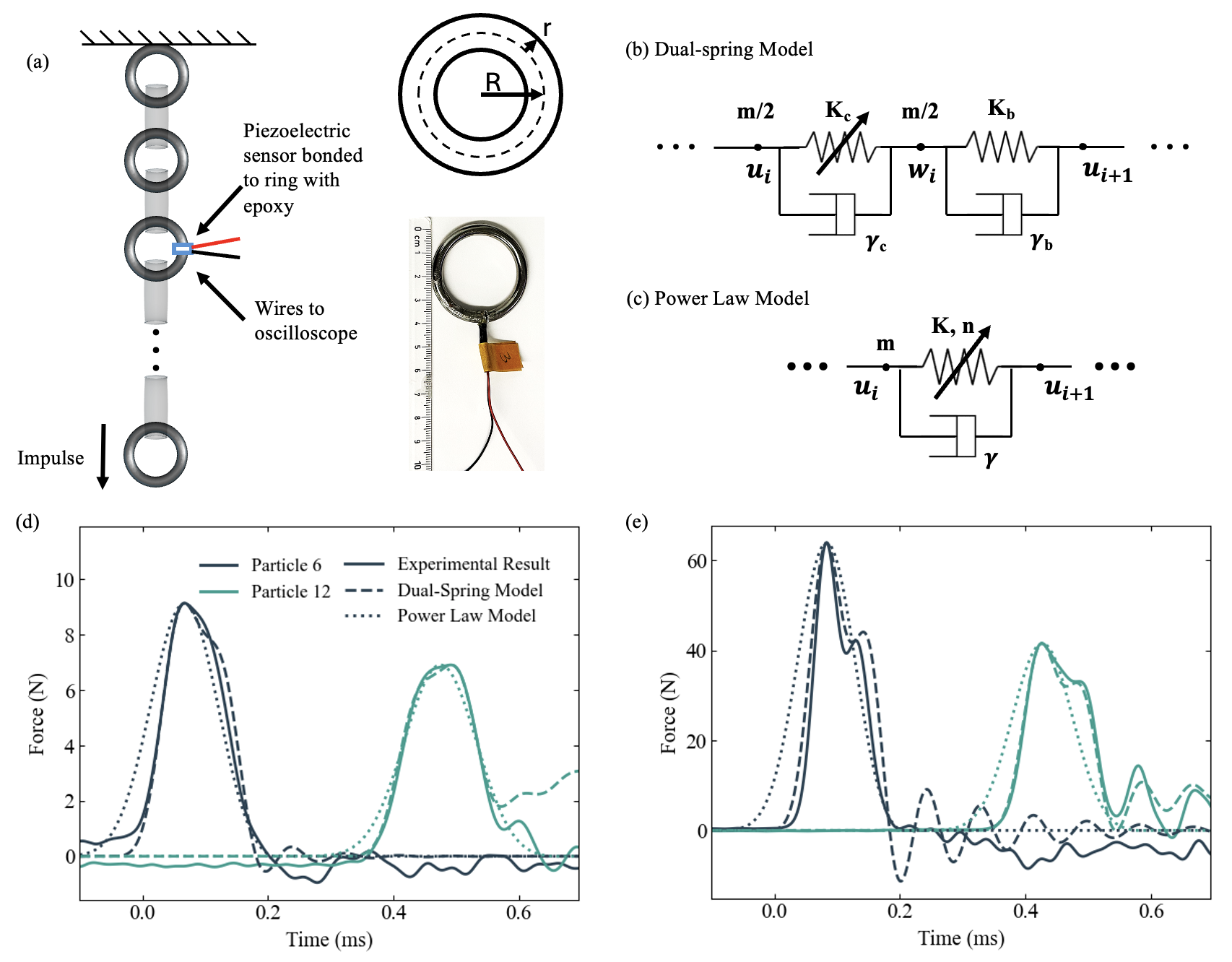}
    \caption{Experimental measurements and numerical modeling of nonlinear wave propagation in a 1D polycatenated ring chain. (a) Experimental setup: a vertical chain of 20 interlocked stainless steel rings. An impulse is generated by dropping the lowermost ring to impact its nearest neighbor, and the resulting wave is captured by piezoelectric sensors embedded in rings 6 and 12 from the impact site. Insets show (top) a schematic of a torus indicating the major radius R and minor radius r, and (bottom) a photograph of a ring with the embedded piezoelectric sensor. (b) Schematic of the hybrid dual-spring unit-cell model. (c) Schematic of the power-law model. (d, e) Comparison of the force-time profiles for a low- (d) and high- (e) amplitude impact. The dual-spring discrete particle simulation (dashed line) and the power-law model (dotted line) are overlaid on the experimental force-time responses (solid line).}
    \label{fig: experiment}
\end{figure*}

\section{\label{sec: exp} Experimental System \& Models}

The system we study is shown in Fig. \ref{fig: experiment}(a): a 
one-dimensional vertical chain of polycatenated tori, gravitationally 
pretensioned by its own weight, in which each ring is interlocked with its 
two nearest neighbors. The inter-ring interaction in a polycatenated chain is governed by a combination of local surface contact and global structural deformation. The contact interface between interlinked tori exhibits a locally saddle-shaped geometry, which can be shown (see Appendix \ref{sec: app contacts}) to follow the Hertzian theory of spherical bodies \cite{Johnson_contact}. The contact force $F_c$ is related to the indentation $\delta_c$ by:
\begin{equation}\label{eq:Fc}
F_c(\delta_c) = \frac{4}{3} E^{\ast}\sqrt{\frac{r(R-r)}{R-2r}} \delta_c^{3/2} = K_c \delta_c^{3/2}, 
\end{equation}
where $E^*$ is the effective elastic modulus, and $R$ and $r$ are the major and minor radii of the torus, as shown in the inset of Fig. \ref{fig: experiment}(a). The ratio $R/r$ thus defines the interlocking geometry and local curvature.

Unlike rigid granular beads, the rings undergo structural bending under load. This contribution to the total compliance is modeled as the deflection of a curved beam. For a ring of major radius $R$, the radial displacement of the ring along the direction of the applied force $F_b$ is expressed as \cite{roark_text}:
\begin{equation}\label{eq:Fb}
    F_b(\delta_b) = \frac{EI}{R^{3}} \left( \frac{\pi k_1}{4} - \frac{2 k_2^2}{\pi} \right)^{-1} \delta_b = K_b \delta_b,
\end{equation}
where $I$ is the cross-sectional second moment of area, $A$ is the cross-sectional area,  $G$ is the shear modulus, and $k_1, k_2$ are constants describing geometry of the system. 

By treating these two mechanisms as two springs in series, as shown in Figure \ref{fig: experiment} (b), the total relative displacement between the centers of adjacent rings is 
\begin{equation}
    \delta= \delta_c(F^{2/3}) + \delta_b(F). \label{eq:dxF}
\end{equation}
We note this static model is similar to the treatment in \cite{zhang2022identification}, which developed a more generic model of contact between round link chains, where the authors also incorporated linear and nonlinear deformation.


To compare the response of this system to the discrete particle framework established for granular crystals \cite{hollow_2013}, we also approximate this hybrid response using a power-law relation (Figure \ref{fig: experiment}(c)):
\begin{equation}
    F(\delta) = K\delta^n. \label{eq:Fdx}
\end{equation}
Here, the effective exponent $n$ serves as a measure of the system's 
overall nonlinearity. The exponent $n$ is either derived from experimental data or, when experimental data are not available, from finite element simulations of the contact mechanics. Having established the contact and bending mechanics 
of a single inter-ring interaction, we next test this model against the 
dynamic response of the full chain (Section~\ref{sec: wave prop}). The 
geometric tunability of $n$ itself is examined separately in 
Section~\ref{sec: tunability}.



\section{\label{sec: wave prop}Nonlinear Wave Propagation in a 1D Ring Chain}

\subsection{Experimental Results}

To investigate the propagation of nonlinear waves in polycatenated systems, we assemble a 1-D vertical chain of 20 stainless steel rings, as shown in Fig. \ref{fig: experiment}(a). Parallel steel rods are positioned to maintain the desired contact orientation between adjacent rings while ensuring no contact with the ring bodies themselves. The chain is subjected to gravitational pretension from its own mass. Impulse is generated by impacting the chain with the lowermost ring, which is raised to varying heights and released. Dynamic force-time responses are captured using two piezoelectric disk sensors (APC International D-5.10mm-.80mm-840) embedded within particles 6 and 12, with force-time signals recorded via a digital oscilloscope (Tektronix TDS 2024). 

Results show the formation of a localized, high-amplitude pulse that propagates through the medium (see Fig. \ref{fig: experiment}(d, e). Notably, the observed waveform morphology consists of a leading solitary wave front followed by a persistent train of trailing oscillations. This differs from the compact, near-symmetric pulses characteristic of classical rigid granular chains, suggesting a more complex internal dynamic within the linked tori.

A hallmark of highly nonlinear systems is the dependence of propagation speed on pulse amplitude. To quantify this in our system, we determined the wave speed $V_s$ by measuring the time-of-flight between peak force arrivals at two distinct sensor locations. Our results demonstrate that as the impact velocity increases, the resulting high-amplitude pulses propagate significantly faster; for example, a low-amplitude pulse reaches the downstream sensor at 0.426 s (Fig. \ref{fig: experiment}d), compared to 0.348s for the high-amplitude counterpart (Fig. \ref{fig: experiment}e).

The trailing oscillations and pronounced amplitude dependence in 
Fig.~\ref{fig: experiment}(d, e) are inconsistent with the compact, 
amplitude-rescaling solitary pulses of a rigid-bead Hertzian chain. Both 
features are, however, consequences of the model established in 
Section~\ref{sec: exp}, where the kinetic energy carried by the leading front is 
partitioned into the rings' internal bending modes, producing a dispersive 
tail superimposed on a nonlinear contact-driven wavefront. To test this 
picture quantitatively, in the following subsection we compare the 
experimental waveforms against discrete-particle simulations of both the 
dual-spring and power-law models introduced in Section~\ref{sec: exp}.

\subsection{Numerical Simulation of Nonlinear Waves}

To investigate the emergence of nonlinear waves from the derived unit-cell mechanics, we model the 1D chain as a system of $N$ unit cells. In the dual-spring model, each unit cell is represented by two distinct mass sites, $u_i$ and $w_i$, each with a mass of $m/2$. The site $u_i$ represents the interface where nonlinear Hertzian contact occurs, while $w_i$ represents the internal node connected via the linear bending spring. The equations of motion for the $i$-th unit cell are given by:

\begin{align}
    \label{eq: dual-spring}
    \frac{m}{2}\ddot{u}_i &= \bigl[F_b(w_{i-1}-u_i) + \gamma_b(\dot{w}_{i-1}-\dot{u}_i)\bigr] \notag \\
&\quad - \bigl[F_c(u_i-w_i) + \gamma_c(\dot{u}_i-\dot{w}_i)\bigr]_{+}, \\[4pt]
    \frac{m}{2}\ddot{w}_i &= \bigl[F_c(u_i-w_i) + \gamma_c(\dot{u}_i-\dot{w}_i)\bigr]_{+} \notag \\
&\quad - \bigl[F_b(w_i-u_{i+1}) + \gamma_b(\dot{w}_i-\dot{u}_{i+1})\bigr].
\end{align}
where $F_c$ represents the unipolar Hertzian contact force and $F_b$ is the bidirectional linear bending force as outlined in equations \ref{eq:Fc} and \ref{eq:Fb}. Here, the subscript $+$ indicates that the contact potential is only activated in tension (corresponding to the physical interlocking of the rings), such that $F_c = 0$ if $\delta < 0$. We numerically integrate these equations using the \texttt{scipy.integrate.odeint} package with a temporal resolution of 10,000 steps per ms to ensure convergence. 

To contextualize our results within the broader field of nonlinear lattice dynamics, we also implement a simplified power-law model as a benchmark. In this framework, the chain is modeled as a sequence of single mass sites $u_i$
(each of mass m) interacting via a lumped potential $F = K\delta^n$, resulting in the following equation of motion:
\begin{align}
\label{eq: power law}
m\ddot{u}_i = &\bigl[F(u_{i-1}-u_i) + \gamma(\dot{u}_{i-1}-\dot{u}_i)\bigr]_{+} \notag \\
 &- \bigl[F(u_i-u_{i+1}) + \gamma(\dot{u}_i-\dot{u}_{i+1})\bigr]_{+}.
\end{align}
We find that this power-law approximation is remarkably effective at reproducing the primary features of the observed signal—specifically the arrival time and pulse width—for low-amplitude impacts where dynamics is dominated by contact interaction. However, the lumped nature of the power-law model inherently lacks the internal degrees of freedom necessary to reproduce the trailing oscillations observed in experiments (Fig. \ref{fig: experiment}(d, e), dotted lines) . 

In contrast, our dual-spring DP simulations (Fig. \ref{fig: experiment}(d, e), dashed lines) successfully capture both the high-amplitude compact front and the subsequent trailing oscillations. This suggests that while the leading pulse carries the chain's translational momentum, a significant portion of the wave energy is partitioned into the ring’s internal bending modes. This is different from classical rigid-sphere chains. The flexibility and geometrical complexity of our polycatenated ring chains introduce additional oscillations onto the solitary wave.

Energy dissipation is incorporated in the dual-spring model via two linear dashpots arranged in series with the linear bending ($\gamma_b$) and nonlinear contact springs ($\gamma_c$), with one dashpot placed in parallel with each spring. The dashpot $\gamma_b$ captures internal structural damping within the ring body, and its coefficient is set by the relative amplitude of the high-frequency trailing oscillations. The second dashpot $\gamma_c$ accounts for losses at the inter-ring contact due to out-of-plane motion and experimental imperfections, and its coefficient is fixed by the amplitude drop between successive pulses. In the power-law model, we lump these two dissipation mechanisms into a single linear dashpot per site ($\gamma$), fitting its damping coefficient with the amplitude of the second peak as the primary constraint.

To align these simulations with observed wave speeds, we replace the particle mass $m$ in Eqs. \ref{eq: dual-spring}-\ref{eq: power law} with an effective mass $m_{eff}$. The effective mass accounts for the portion of the ring’s mass actively participating in longitudinal translational momentum; the remaining inertia is localized in transverse ovalization and bending, and does not contribute to the longitudinal wave speed. The use of effective mass is also found in literature on  granular columns \cite{liu2009size, hsu2009dynamic}, where internal degrees of freedom decouple from the primary wavefront, and the $m_{\text{eff}}$ values for our polycatenated chains (Tables \ref{Tab: ratio exp}, \ref{Tab: angle exp}) fall within ranges reported for these systems.

\section{\label{sec: tunability}Tunable Nonlinearity}

While the dual-spring model provides a more complete physical description by capturing the energy partitioning into internal bending modes, the power-law model offers a simplified  framework that effectively characterizes the primary solitary wave front. This abstraction is particularly powerful because it reduces the complex, multi-modal interaction into a single effective nonlinear exponent, $n$, which serves as a direct measure of the system's overall nonlinearity. By adopting this power-law representation, we can treat the degree of nonlinearity not as a fixed material constant—as is the case with the $n=1.5$ exponent in classical Hertzian spheres—but as a designable geometric property. We now explore how the specific geometry of the unit cell dictates the resulting wave dynamics in polycatenated ring chains.

\subsection{Tuning Nonlinearity via Ring Aspect Ratio $R/r$}

We performed finite element simulations in Abaqus/CAE using the implicit dynamics solver to study the force-displacement relationship between a pair of half rings in contact. In the simulation, we fix the end of one half ring by boundary conditions, and apply a controlled displacement at the end of the other half ring. We define a surface-to-surface hard contact interaction between the rings, and measure the resulting force at the fixed boundary. 

\begin{figure}[ht]
    \centering
    \includegraphics[width=\linewidth]{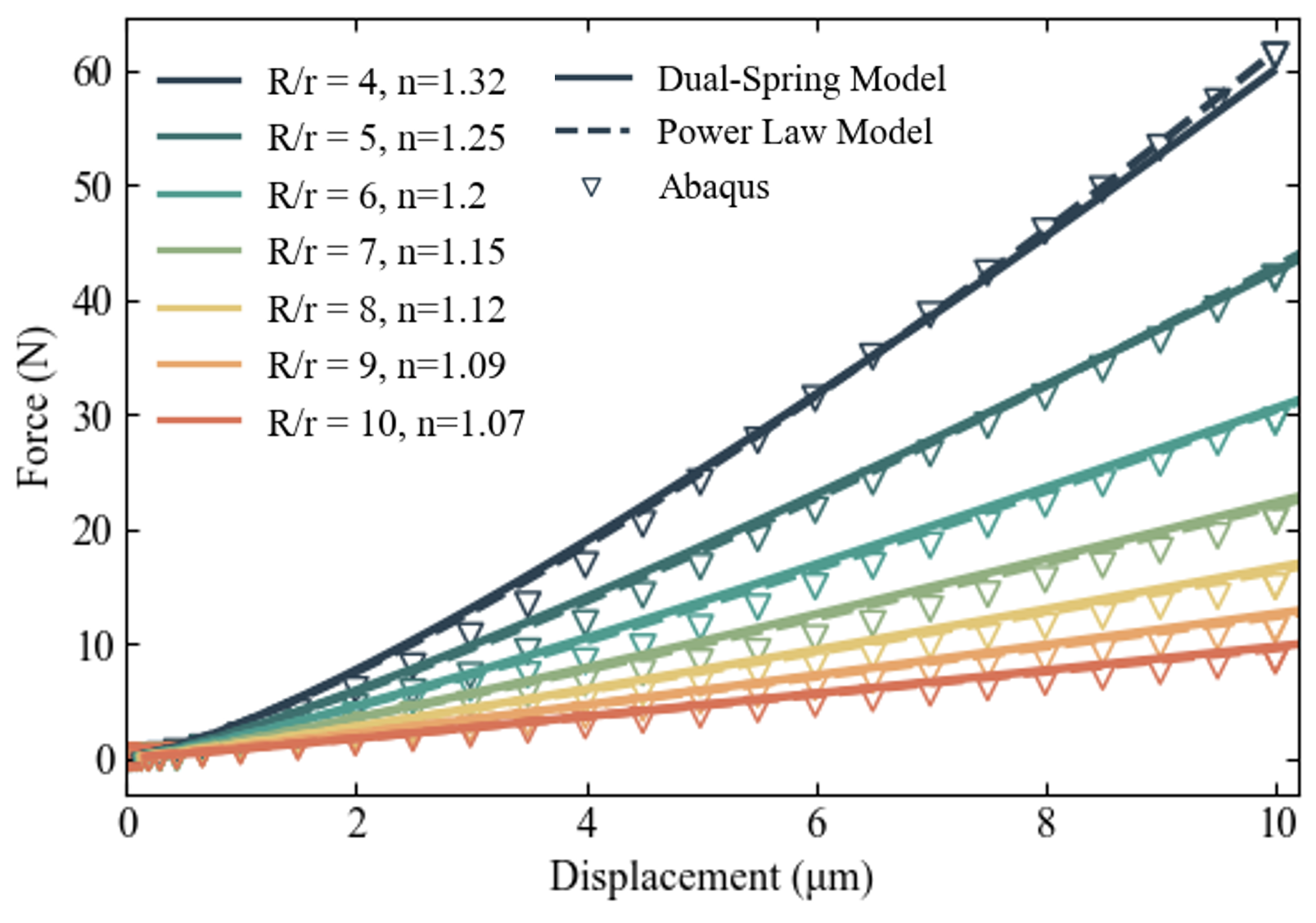}
    \caption{Study of the effect of ring aspect ratio R/r on the nonlinear contact exponent n. We simulate the contact between two rings with cross section radius r=2.5mm and vary the ring body radius R. For small forces and displacements, the contact theory with two springs (solid lines) and its power law fit (dashed lines) are consistent with finite element simulation in Abaqus (triangles).}
    \label{fig:theory_ratios}
\end{figure}

In Figure \ref{fig:theory_ratios}, we compare these simulation results to the theoretical force-displacement relationships predicted by equation \ref{eq:dxF} and \ref{eq:Fdx} for different physically admissible aspect ratios $R/r>3$. Assuming a constant cross section radius r, the fitted power law exponent $n$ varies from 1.07 to 1.32 as the aspect ratio $R/r$ decreases from 10 to 4. This range of $n$ spans from near-linear behavior ($n \to 1$) to more pronounced nonlinear response. For high aspect ratios, the smaller contact exponent indicates that the contact interaction is dominated by the elastic ring deflection in thinner rings. As the aspect ratio decreases, the contribution of the Hertzian contact deformation becomes more significant, leading to an increase in $n$. At the lowest aspect ratio allowed geometrically ($R/r = 3.1$), we find $n = 1.418$, approaching the Hertzian contact exponent of 1.5.

\subsection{Tuning Nonlinearity via Contact Angles}

In previous studies of granular systems composed of cylindrical elements, the contact angle between neighboring particles has been studied as a key parameter that influences the contact stiffness and nonlinearity \cite{li2012tunable}. Thus we expect the nonlinearity to be further modulated by the contact angle $\theta$ between adjacent rings.  Although the theory model discussed in Appendix \ref{sec: app contacts}. is derived for normal contact ($\theta =90^\circ$), it can be generalized for oblique contacts ($\theta =45^\circ$), where the contact area is an ellipse \cite{wang2013hertz, garland2011elliptical}. However, at larger misalignments $\theta < 45^\circ$, two regions of contact emerge between neighboring ring pairs, which is inconsistent with the single-ellipse Hertzian assumption. In these regimes, we proceed to study these systems in Abaqus finite-element simulation.

\begin{figure}[t]
    \centering
    \includegraphics[width=\linewidth]{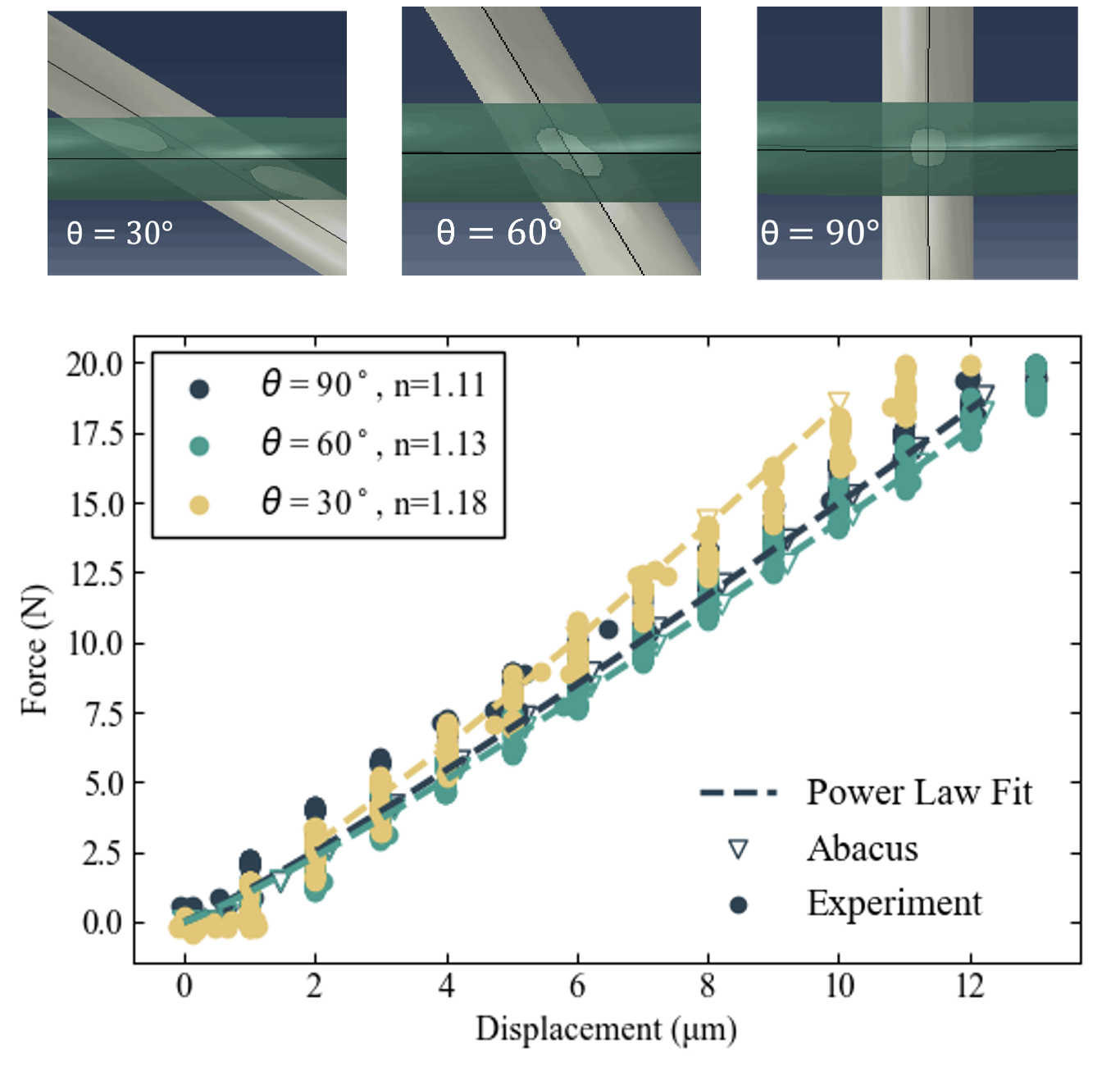}
    \caption{Study of the effect of ring contact angle $\theta$ on the nonlinear contact exponent n. Top: two rings in contact and the shape of their contact (shaded regions) for different angles $\theta = 90^\circ, 60^\circ, 30^\circ$. Bottom: Comparison of the force-displacement relationship between two rings with aspect ratio $R/r=9$ in contact at different angles. For small forces and displacements, the experimental results (solid circles) are consistent with finite element simulation in Abaqus (triangles).}
    \label{fig:exp_angles}
\end{figure}

In Fig. \ref{fig:exp_angles}, we compare the force-displacement relations of a pair of steel rings ($R = 20\textrm{mm}$, $r=2.5\textrm{mm}$) in contact at various angles. Both the Abaqus simulation and quasi-static tensile tests reveal that the fitted contact exponent $n$ increases as the twist angle $\theta$ decreases, indicating stronger nonlinearity for an oblique contact. Notably, this nonlinearity persists even in the regime where $\theta <45^\circ$. The similarity of contact dynamics also suggests that analogous dynamical responses, such as formation and propagation of solitons, may continue to be supported in twisted ring chains, as will be observed in Sec. \ref{sec: dynamic}. Further, variations in contact angle due to vibrations near normal contact in experiment are unlikely to destroy the solitary wave dynamics.

\begin{table*}[t]
    \centering
    \includegraphics[width=0.95\textwidth]{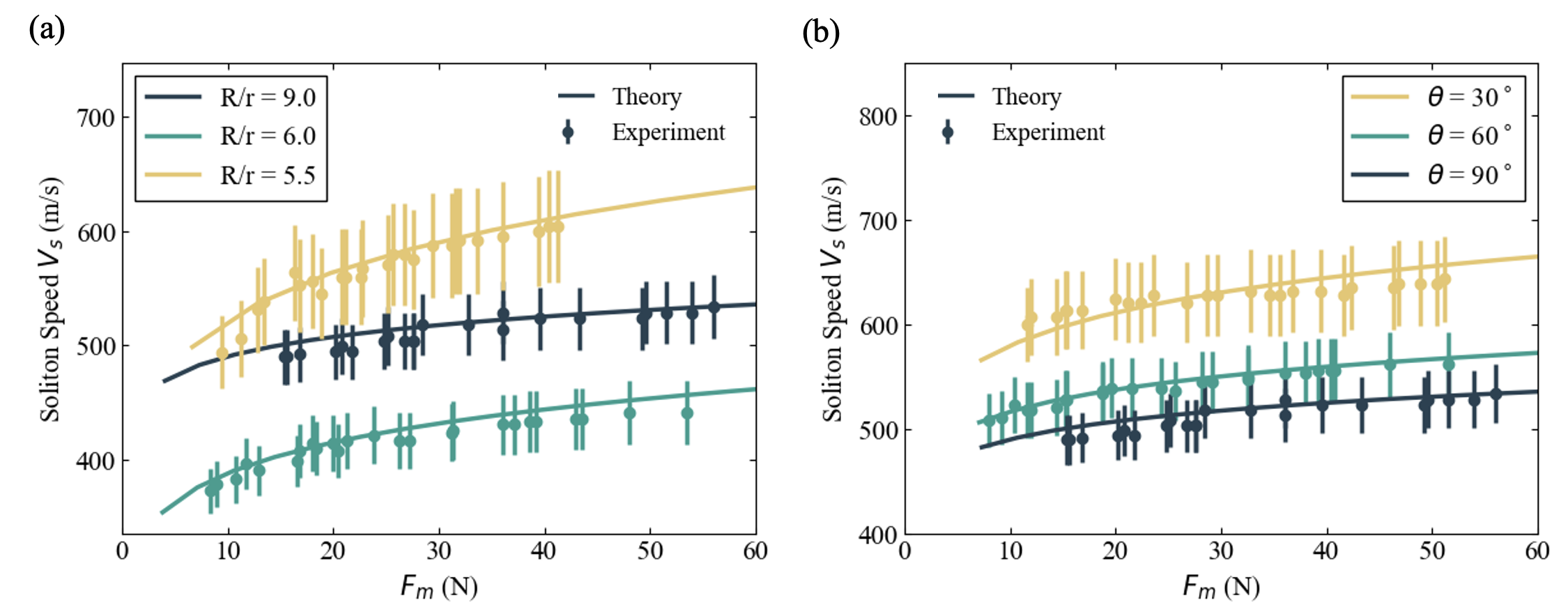}
    \captionof{figure}{Comparison of experimental measurements (solid circle) of solitary wave speed $V_s$ with theory predictions from Eqn. \ref{eq:vsfm} (solid lines) for various ring geometric ratios (a) and contact angles (b).}
    \label{fig: tunability}
    
    \vspace{2em} 

    \begin{minipage}{0.48\textwidth}
        \centering
        \captionof{table}{Geometric and dynamic parameters of ring-chain samples with different aspect ratios $R/r$. $K$ and $n$ are from quasi-static tests; $m_{\text{eff}}/m_0$ is from dynamic fitting.}
        \label{Tab: ratio exp}
        \vspace{0.5em}
        \renewcommand{\arraystretch}{1.2}
        \setlength{\tabcolsep}{6pt}
        \begin{tabular}{cccc}
        \hline
        \makecell{\textbf{Aspect}\\\textbf{Ratio} \\ $R/r$} &
        \makecell{\textbf{Spring} \\\textbf{Constant}\\$K$ (N/mm$^{n}$)} &
        \makecell{\textbf{Contact}\\\textbf{Exponent}\\ $n$} &
        \makecell{\textbf{Effective}\\\textbf{Mass}\\ $m_{\text{eff}}/m_0$} \\
        \hline
        9.0  & 2284    & 1.11 & 0.37 \\
        6.0  & 5044    & 1.24 & 0.35 \\
        5.5  & 17700   & 1.29 & 0.35 \\
        \hline
        \end{tabular}
    \end{minipage}
    \hfill 
    \begin{minipage}{0.48\textwidth}
        \centering
        \captionof{table}{Summary of contact parameters for different contact angles $\theta$. $K$ and $n$ are from quasi-static tests; $m_{\text{eff}}/m_0$ is from dynamic fitting.}
        \label{Tab: angle exp}
        \vspace{0.5em}
        \renewcommand{\arraystretch}{1.2}
        \setlength{\tabcolsep}{6pt}
        \begin{tabular}{cccc}
        \hline
        \makecell{\textbf{Contact}\\\textbf{Angle}\\ $\theta$} &
        \makecell{\textbf{Spring} \\ \textbf{Constant}\\$K$ (N/mm$^{n}$)} &
        \makecell{\textbf{Contact}\\\textbf{Exponent} \\$n$} &
        \makecell{\textbf{Effective}\\\textbf{Mass} \\ $m_{\text{eff}}/m_0$} \\
        \hline
        $90^{\circ}$ & 2284  & 1.11 & 0.37 \\
        $60^{\circ}$ & 2605  & 1.13 & 0.34 \\
        $30^{\circ}$ & 4030  & 1.18 & 0.33 \\
        \hline
        \end{tabular}
    \end{minipage}
\end{table*}

\subsection{Dynamical Tunability}
\label{sec: dynamic}

By reducing the complex hybrid interaction to a nonlinear exponent $n$, we can map the system onto the analytical framework established by Nesterenko for generalized power-law media \cite{Nesterenko-book-2001}. This abstraction allows us to explore the tunability of the chain's nonlinearity, specifically how the wave speed scales with amplitude, as a function of the geometric parameters $R/r$ and $\theta$.

For systems characterized by a generalized power-law contact interaction, Nesterenko derived analytical solutions using a continuous model and the long-wavelength approximation. These results can be adapted to our polycatenated system (detailed in Appendix A) to describe the propagation of compact solitary waves. The resulting wave speed $V_s$ depends nonlinearly on the maximum dynamic force $F_m$ according to:
\begin{equation}
\label{eq:vsfm}V_s = a\sqrt{\frac{2}{(n+1)}} \biggl(\frac{K^{1/n}}{m}\biggr)^{1/2} F_m^{(n-1)/2n},
\end{equation}
where $a = 2(R-r)$ represents the initial distance between particle centers and $m$ is the particle mass.

In polycatenated ring chains, the degree of nonlinearity is not a fixed material constant but a tunable geometric property. By varying the aspect ratio $R/r$ and the inter-ring contact angle $\theta$, we can shift the system response between a bending-dominated regime ($n \to 1$) and a contact-dominated regime ($n \to 1.5$). Plotting the observed solitary wavefront speeds against Nesterenko’s theory across this geometric parameter space (Figure \ref{fig: tunability}) confirms that the primary pulse dynamics remain robustly governed by the effective exponent $n$. 

However, the persistence of trailing oscillations indicates a significant partitioning of energy into internal degrees of freedom. The structural bending modes (modeled as springs $K_b$) act as local sinks for energy, which necessitates a renormalization of the mass participating in the translational wave dynamics through an effective mass $m_{\text{eff}}$. As shown in the wave speed comparisons in Fig. \ref{fig: tunability}, the experimental data and theory curves align well, with the system parameters and fitted $m_{\text{eff}}$ presented in Table \ref{Tab: ratio exp} and \ref{Tab: angle exp}.

\section{\label{sec: discussion}conclusion}

We report the first experimental and numerical investigation of nonlinear solitary wave propagation in one-dimensional chains of polycatenated rings. By characterizing the inter-ring interaction as a coupled mechanism of Hertzian contact and structural bending, we demonstrate that these topologically interlocked, tension-activated systems support amplitude-dependent nonlinear wave propagation. This modeling is validated through finite-element analysis, discrete particle simulations, and dynamic sensing experiments. Notably, unlike classical rigid-sphere granular crystals, polycatenated chains exhibit a unique wave morphology: a leading wavefront followed by distinct trailing oscillations. This phenomenon arises from energy partitioning into internal bending modes, providing an intrinsic dispersive mechanism that fundamentally distinguishes these highly deformable architected materials from ideal rigid-body granular systems.

Crucially, we demonstrate that the system's effective nonlinearity is not a fixed material constraint, but a tunable feature. The wave dynamics are governed by a nonlinear power-law exponent, which can be controlled by altering geometric parameters such as the ring aspect ratio $R/r$ and contact angle $\theta$. This design freedom shifts strongly nonlinear wave properties from static material constants to designable structural features, establishing polycatenated chains as a powerful new platform for dynamic and tunable control of nonlinear waves. Future research may expand this foundational 1D model to two-dimensional interlinked lattices \cite{leonard_traveling_2014, hua_wave_2019}, and further explore the effects of inherent dissipation and disorder on wave propagation.

\section*{Data Availability}

The data that support the findings of this study are available from the corresponding author upon reasonable request.

\begin{acknowledgments}
C. D. and X. X. acknowledge funding from the New Frontiers of Sound Center: NSF Award No. 2242925. R.Y. is grateful for the support provided by the Funai Overseas Scholarship. The authors are grateful to Jihoon Ahn, Mehmet Naci Keskin, Sujeeka Nadarajah, Max Tepermeister, and Hujie (Henry) Yan for insightful conversations, training, and experimental support.
\end{acknowledgments}

\bibliography{references}
\appendix

\section{\label{sec: app contacts}Theory of Nonlinear Contact in Rings}
To analyze this interaction, we begin with the Hertz contact theory\cite{Hertz_original_1882}, which first examines the separation between two surfaces in contact. This separation $h$ is described by the following equation\cite{Johnson_contact}:
\begin{equation}\label{eq:one}
h = A x^2 + B y^2 = \frac{1}{2R'} x^2 + \frac{1}{2R''} y^2.
\end{equation}
The parameters $A$ and $B$ are defined by the following equations:
\begin{subequations}
\label{eq:whole}
\begin{equation}
    (A+B) = \frac{1}{2} \left( \frac{1}{R_1'} + \frac{1}{R_1''} + \frac{1}{R_2'} + \frac{1}{R_2''}\right), \label{subeq:1}
\end{equation}
and 
\begin{eqnarray}
        |A-B| &&= \frac{1}{2} \Biggl\{ \left( \frac{1}{R_1'} - \frac{1}{R_1''}\right)^2 + \left( \frac{1}{R_2'} - \frac{1}{R_2''}\right)^2 \nonumber\\
        &&\quad + 2\left( \frac{1}{R_1'} - \frac{1}{R_1''}\right)\left( \frac{1}{R_2'} - \frac{1}{R_2''}\right) \cos{2\alpha} \Biggl\}^{1/2},
        \label{subeq:2}
\end{eqnarray}
\end{subequations}
where, $R_1'$ and $R_1''$ are the principal radii of curvature of body 1 at the contact point, $R_2'$ and $R_2''$ are the corresponding principal radii of curvature of body 2, respectively, and $\alpha$ represents the angle of inclination between the principal axes of curvature of the two bodies.
For the perpendicular orientation of tori, which we consider in this paper, $\alpha = \pi/2$. Assuming the rings have identical dimensions, substituting $R_1' = R_2' = r$ and $R_1''=R_2''=-(R-r)$, we get $R' = R'' = \frac{r(R-r)}{R-2r}$, where $R$ is the major radius and $r$ is the minor radius of the torus. This expression assumes $R>2r$, so that the local relative curvature is positive and the Hertz contact coefficient is real. It implies that the contact area of the tori is circular, and consequently, the contact force follows the Hertz theory of spherical bodies\cite{Johnson_contact}:
\begin{equation}\label{eq:three}
    F = k_c \delta ^{3/2} = \frac{4}{3} E^{\ast}\sqrt{\frac{r(R-r)}{R-2r}} \delta ^{3/2}
\end{equation}
where, $E^{\ast}$ is the effective elastic modulus defined as $1/E^{\ast} = (1-\nu_1^2)/E_1 + (1-\nu_2^2)/E_2$, where $E_1$, $E_2$ and $\nu_1$, $\nu_2$ are the Young's moduli and Poisson's ratios of the two bodies, respectively. The parameter $\delta$ represents the normal approach, or overlap, between the two bodies. It is worth noting that for non-perpendicular orientations of the tori ($\alpha \neq \pi/2$), the contact area would be elliptical, necessitating the application of the Hertz theory for elliptical contacts.

\section{Theory of Nonlinear Wave Propagation}
Here, we derive the analytical solution of the solitary wave on the interlocked torus rings, with a generalized power law $n>1$. We will follow the method established by Nesterenko\cite{Nesterenko-1992,Nesterenko_1994, Nesterenko-book-2001}, to determine the solitary wave speed $V_s$ and the typical length scale of the solitary wave $L$ in our system.

We start by treating the torus chain as a continuous medium, where the displacement at bead $i$ can be written as $u_i = u(x, t)$, with $x$ being the spatial coordinate along the chain and $t$ being time. We assume that $L$ is larger than the characteristic distance $a = 2(R-r)$ between the contact points of the torus rings. This approximation is known as the long-wavelength approximation. The displacement at neighboring beads can be expanded as
\begin{align}\label{eq:disp_expansion}
    u_i &= u, \\
    u_{i-1} &= u - au_x + \frac{a^2}{2}u_{xx} - \frac{a^3}{6}u_{xxx} + \frac{a^4}{24}u_{xxxx} + \cdots, \\
    u_{i+1} &= u + au_x + \frac{a^2}{2}u_{xx} + \frac{a^3}{6}u_{xxx} + \frac{a^4}{24}u_{xxxx} + \cdots.
\end{align}  
We can now express the displacement differences as
\begin{align}
    u_{i+1} - u_i &= N + \varphi,\\
    u_i - u_{i-1} &= N + \psi,
\end{align}
where,
\begin{align*}
    N &= a u_x,\\
    \varphi &= \frac{a^2}{2} u_{xx} + \frac{a^3}{6} u_{xxx} + \frac{a^4}{24} u_{xxxx} + \cdots,\\
    \psi &= -\frac{a^2}{2} u_{xx} + \frac{a^3}{6} u_{xxx} - \frac{a^4}{24} u_{xxxx} + \cdots.
\end{align*}
The interaction terms in the equation of motion can be approximated as,
\begin{align}\label{6}
    (u_{i+1} - u_i)_{+}^n &\approx (N + \varphi)_{+}^n,\\
    \label{7}
    (u_{i} - u_{i-1})_{+}^n &\approx (N + \psi)_{+}^n.
\end{align}
As N and $\varphi$, $\psi$ have different order of magnitude with respect to the small parameter $\varepsilon = a/L$,
\begin{equation}
    \frac{\varphi}{N} \sim \frac{\psi}{N} \sim \varepsilon,
\end{equation}
which we can use to expand the expressions on the right-hand side of Equations (\ref{6}) and (\ref{7}) as
\begin{equation}
    \begin{split}
        (N + \varphi)_{+}^n &= N^n + nN^{n-1}\varphi + \frac{n(n-1)}{2} N^{n-2} \varphi^2 \\
        &+ \frac{n(n-1)(n-2)}{6} N^{n-3} \varphi^3 + \cdots,
    \end{split}
\end{equation}
\begin{equation}
    \begin{split}
   (N + \psi)_{+}^n &= N^n + nN^{n-1}\psi + \frac{n(n-1)}{2} N^{n-2} \psi^2 \\
   &+ \frac{n(n-1)(n-2)}{6} N^{n-3} \psi^3 + \cdots. 
    \end{split}
\end{equation}
By plugging these two expansions into Eqs. \ref{6} and \ref{7} and re-arranging the terms, the equation of motion for the displacement \( u(x, t) \) can now be expressed as a partial differential equation of the form
\begin{align}
    u_{tt} &= AnN^{n-1} (\varphi - \psi) + A \frac{n(n-1)}{2} N^{n-2} (\varphi^2 - \psi^2) \notag \\
    &\quad + A \frac{n(n-1)(n-2)}{6} N^{n-3} (\varphi^3 - \psi^3) + \cdots,
\end{align}
with $A=K/m$.

Using the definitions of $\varphi$ and $\psi$, one can find
\begin{align}
    \varphi - \psi &= a^2 u_{xx}+  \frac{a^4}{12} u_{xxxx} + \mathcal{O}(a^6),\\
    \varphi^2 - \psi^2 &= \frac{a^5}{3} u_{xx} u_{xxx} + \mathcal{O}(a^7), \\
    \varphi^3 - \psi^3 &= \frac{a^6}{4} u_{xx}^3 + \mathcal{O}(a^8),
\end{align}
which implies
\begin{equation}\label{15}
    \begin{split}
        u_{tt} = Aa^{n+1} [&n u_x^{n-1} u_{xx} + \frac{n}{12} a^2 u_x^{n-1} u_{xxxx} \\
        &+ \frac{n(n-1)}{6} a^2 u_x^{n-2} u_{xx} u_{xxx}\\
        &+ \frac{n(n-1)(n-2)}{24} a^2 u_x^{n-3} u_{xx}^3] ,
    \end{split}
\end{equation}
where \( N = au_x \) was used. The retained terms are accurate through relative order $\mathcal{O}(\varepsilon^2)$ in the long-wavelength expansion; higher-order corrections are neglected. Equation (\ref{15}) can be written in a more convenient form as
\begin{equation}
    \begin{split}
        u_{tt} &= Aa^{n+1} \left\{ u_x^n + \frac{na^2}{24} \left[ (n-1)u_x^{n-2} u_{xx}^2 + 2u_x^{n-1} u_{xxx} \right] \right\}_x.
    \end{split}
\end{equation}

for \( u_x > 0 \) (tension). Defining \( \xi := u_x \) and differentiating in \( x \), we arrive at the governing equation:
\begin{equation}\label{17}
    \xi_{tt} = c_n^2 \left\{ \xi^n + \frac{na^2}{24} \left[ (n-1) \xi^{n-2} \xi_x^2 + 2 \xi^{n-1} \xi_{xx} \right] \right\}_{xx},
\end{equation}
where \( c_n^2 = Aa^{n+1} \). Note that the $\xi$ is defined with the positive sign, as opposed to the conventional granular chain. $\xi > 0$ ensures the system is under the tension. 

Since solitary waves are translated along the chain without changing the waveform, the solution is expected to be stationary, a function of $\tau := x - V_st$, where $V_s$ is the wave speed of the solitary wave. Using the chain rule
\begin{equation}\label{18}
    \frac{\partial \xi}{\partial x} = \frac{\partial \xi}{\partial \tau} \frac{\partial \tau}{\partial x} = \frac{\partial \xi}{\partial \tau} \quad \text{and} \quad \frac{\partial \xi}{\partial t} = \frac{\partial \xi}{\partial \tau} \frac{\partial \tau}{\partial t} = -V_s \frac{\partial \xi}{\partial \tau},
\end{equation}
Equation (\ref{17}) can be expressed as an ordinary differential equation, which upon integrating twice in \( \tau \) yields
\begin{equation}\label{19}
    \left( \frac{V_s}{c_n} \right)^2 \xi = \xi^n + \frac{na^2}{24} \left[ (n-1) \xi^{n-2} \xi_{\tau}^2 + 2 \xi^{n-1} \xi_{\tau \tau} \right] + C_1 \tau + C_0,
\end{equation}
with $C_0$ and $C_1$ being integration constants. As $\xi$ needs to stay finite for $\tau \rightarrow \pm \infty$, $C_1$ needs to vanish. Furthermore, for a localized solitary wave with $\xi, \xi_\tau, \xi_{\tau\tau} \rightarrow 0$ as $|\tau| \rightarrow \infty$, we also have $C_0 = 0$.

Introducing the change of variables
\begin{equation}
    y = \xi^{\frac{n+1}{2}} \left( \frac{c_n}{V_s} \right)^{\frac{n+1}{n-1}}, \quad \mu = \frac{\tau}{a} \sqrt{\frac{6(n+1)}{n}},
\end{equation}
(\ref{19}) reduces to the equation of motion for a nonlinear oscillator
\begin{equation}\label{21}
    y_{\mu\mu} = -\frac{\partial W(y)}{\partial y},
\end{equation}
where \( W \) is the oscillator potential field
\begin{equation}
    W(y) = \frac{y^2}{2} - \frac{n+1}{4} y^{\frac{4}{n+1}} ,
\end{equation}
where we have used $C_0=0$ for the localized solitary wave (a non-zero $C_0$ would generate an additional $C y^{2/(n+1)}$ term in $W(y)$, with $C$ a constant determined by $C_0$).

One solution of (\ref{21}) is
\begin{equation}
    \xi = \left\{ \frac{(n+1)}{2} \frac{V_s^2}{c_n^2} \right\}^{\frac{1}{n-1}} \left| \sin \left( \frac{(n-1)}{(n+1)} \sqrt{\frac{6(n+1)}{n}} \frac{\tau}{a} \right) \right|^{\frac{2}{n-1}}.
\end{equation}

The compact solitary-wave solution is obtained by taking one hump of the sine profile, over a span $\pi/\omega$, and setting $\xi=0$ outside that interval. At the peak of the solitary wave, the sine-term is equal to one and we can solve for the solitary wave velocity to find
\begin{equation}
        V_s = \sqrt{\frac{2c_{n}^2}{n+1} \xi_{\text{max}}^{n-1}} = \left( \frac{2c_{n}^2}{n+1} \right)^{\frac{1}{n+1}} (-v_{\text{max}})^{\frac{n-1}{n+1}},
\end{equation}
where $v_{\text{max}}$ is the signed peak bead velocity in the solitary wave. We can get the relationship $\xi_{\text{max}} = -v_{\text{max}}/V_s$ by using the chain rule (\ref{18}).

To express the wave speed in terms of the maximum dynamic force, we use
\begin{equation}
    F_m = K\delta_{\max}^n, 
    \qquad 
    \delta_{\max} \simeq a\xi_{\max}.
\end{equation}
Thus,
\begin{equation}
    \xi_{\max}
    =
    \frac{1}{a}\left(\frac{F_m}{K}\right)^{1/n}.
\end{equation}
Substituting this into
\begin{equation}
    V_s^2 = \frac{2c_n^2}{n+1}\xi_{\max}^{n-1},
\end{equation}
and using $c_n^2=Ka^{n+1}/m$, gives
\begin{equation}
    V_s = a\sqrt{\frac{2}{n+1}} \left(\frac{K^{1/n}}{m}\right)^{1/2}
    F_m^{\frac{n-1}{2n}}.
\end{equation}
\end{document}